\documentclass[twocolumn,preprintnumbers,prl,nofootinbib]{revtex4-1}

\usepackage{amsmath,amssymb,slashed}
\usepackage{hyperref}
\usepackage{url}
\usepackage{breakurl}
\usepackage{graphicx}

\def\bra#1{\langle #1 |}
\def\ket#1{| #1 \rangle}

\def\app#1#2{%
  \mathrel{%
    \setbox0=\hbox{$#1\sim$}%
    \setbox2=\hbox{%
      \rlap{\hbox{$#1\propto$}}%
      \lower1.1\ht0\box0%
    }%
    \raise0.25\ht2\box2%
  }%
}

\begin{document}

\preprint{UCB-PTH-14/42}

\title{The Black Hole Interior in Quantum Gravity}

\author{Yasunori Nomura, Fabio Sanches, and Sean J. Weinberg}
\affiliation{Berkeley Center for Theoretical Physics, Department of Physics, 
  University of California, Berkeley, CA 94720}
\affiliation{Theoretical Physics Group, Lawrence Berkeley National 
  Laboratory, Berkeley, CA 94720}

\begin{abstract}
We discuss the interior of a black hole in quantum gravity, in which 
black holes form and evaporate unitarily.  The interior spacetime appears 
in the sense of complementarity because of special features revealed 
by the microscopic degrees of freedom when viewed from a semiclassical 
standpoint.  The relation between quantum mechanics and the equivalence 
principle is subtle, but they are still consistent.
\end{abstract}

\maketitle

\section{Introduction}

Despite much effort, the relation between quantum mechanics and the 
spacetime picture of general relativity has never been clear.  The issue 
becomes particularly prominent in black hole physics~\cite{Preskill:1992tc}. 
Quantum mechanics suggests that the black hole formation and evaporation 
processes are unitary---a black hole simply appears as an intermediate 
resonance between the initial collapsing matter and final Hawking 
radiation states~\cite{'tHooft:1990fr}.  Meanwhile, general relativity 
suggests that an observer falling into a large black hole does not 
feel anything special at the horizon.  These two assertions are 
surprisingly hard to reconcile.  With naive applications of quantum 
field theory on curved spacetime, one is led to the conclusion that 
unitarity of quantum mechanics is violated~\cite{Hawking:1976ra} or 
an infalling observer finds something dramatic (a firewall) at the 
horizon~\cite{Almheiri:2012rt,Almheiri:2013hfa,Marolf:2013dba,%
Braunstein:2009my}.

In this letter, we argue that the resolution to this puzzle lies in 
how a semiclassical description of the system arises from the microscopic 
theory of quantum gravity.  While a semiclassical description employs 
an {\it exact} spacetime background, the quantum uncertainty principle 
implies that there is no such thing---there is an intrinsic uncertainty 
for background spacetime for any finite energy and momentum.  This 
implies that at the microscopic level there are many different ways 
to arrive at the same background for the semiclassical theory, within 
the precision allowed by quantum mechanics.  This is the origin of 
the Bekenstein-Hawking entropy~\cite{Bekenstein:1973ur,Hawking:1974sw}. 
The semiclassical picture is obtained after coarse-graining 
these degrees of freedom, which we call {\it vacuum degrees of 
freedom}~\cite{Nomura:2014yka}.

We argue that much of the puzzle regarding unitary evolution and the 
interior spacetime of a black hole arises from peculiar features the 
vacuum degrees of freedom exhibit when viewed from the semiclassical 
standpoint.  In particular, they show properties which we call 
{\it extreme relativeness} and {\it spacetime-matter duality}. 
The first refers to the fact that the spacetime distribution of these 
degrees of freedom changes when we adopt a different ``reference 
frame.''  This change occurs in a way that the answers to any 
physical question are consistent with each other when asked 
in different reference frames. Together with the reference frame 
dependence of the semiclassical degrees of freedom discussed 
earlier~\cite{Susskind:1993if,Susskind:2005js}, this comprises basic 
features of how general coordinate transformations work in the full 
theory of quantum gravity.

The second property is related to the following fact:\ while the 
vacuum degrees of freedom are interpreted as how the semiclassical 
spacetime is realized at the microscopic level, their interactions 
with semiclassical degrees of freedom make them look like thermal 
radiation.  In fact, these degrees of freedom are neither spacetime 
nor matter/radiation, as indicated by the fact that their spacetime 
distribution is frame dependent, and that their detailed dynamics 
cannot be treated in semiclassical theory.  This situation reminds 
us of wave-particle duality---a quantum object exhibits dual properties 
of waves and particles while the ``true'' (quantum) description does 
not fundamentally rely on either of these classical concepts.

The two properties described above allow us to avoid the arguments 
in Refs.~\cite{Almheiri:2012rt,Almheiri:2013hfa,Marolf:2013dba} and 
make the existence of the black hole interior consistent with unitary 
evolution, in the sense of complementarity~\cite{Susskind:1993if} 
as envisioned in Refs.~\cite{Nomura:2011dt,Nomura:2013nya}.  A notion 
of geometry carrying information has also been considered recently 
in Ref.~\cite{Perez:2014xca} in a different model of black hole 
evolution; see also Ref.~\cite{Rovelli:1996dv} for early discussions. 
In our picture, we assume that a black hole evaporates through 
Hawking radiation~\cite{Hawking:1974sw}; for an alternative view, 
see Ref.~\cite{Haggard:2014rza}.

In the rest of the letter, we present our picture using the example of 
a Schwarzschild black hole formed by collapsing matter in 4-dimensional 
spacetime.  More detailed descriptions are given in the accompanying 
paper~\cite{NSW}.

\section{Distant Description}

Consider a quantum state representing a black hole of mass $M$ located 
at some place at rest, as described in a distant reference frame.  (We 
adopt the Schr\"{o}dinger picture throughout.)  Because of the uncertainty 
principle, such a state must involve a superposition of energy and 
momentum eigenstates.  In particular, since a black hole of mass $M$ 
will evolve after Schwarzschild time $\varDelta t \approx O(M l_{\rm P}^2)$ 
into a state representing a Hawking quantum and a smaller mass black 
hole, the state must involve a superposition with
\begin{equation}
  \varDelta E \approx \frac{1}{\varDelta t} 
    \approx O\biggl(\frac{1}{M l_{\rm P}^2}\biggr),
\label{eq:delta-E}
\end{equation}
where $E$ is defined in the asymptotic region, and $l_{\rm P}$ the 
Planck length.  Requiring that the position uncertainty is comparable 
to the quantum stretching of the horizon $\varDelta r \approx O(1/M)$, 
where $r$ is the Schwarzschild radial coordinate, the momentum spread 
is $\varDelta p \approx O(1/M l_{\rm P}^2)$.  This gives an uncertainty 
of the kinetic energy much smaller than $\varDelta E$, so the spread 
of the energy comes mostly from a superposition of different rest 
masses:\ $\varDelta E \approx \varDelta M$.

How many different independent ways are there to superpose the 
energy eigenstates to arrive at the same black hole geometry within 
this precision?  We assume that the Bekenstein-Hawking entropy, 
${\cal A}/4 l_{\rm P}^2$, gives the logarithm of this number (at 
the leading order in $l_{\rm P}^2/{\cal A}$), where ${\cal A} = 
16\pi M^2 l_{\rm P}^4$ is the area of the horizon.  The nonzero 
Bekenstein-Hawking entropy implies that there are exponentially 
many independent black hole {\it vacuum} states in a small energy 
interval of Eq.~(\ref{eq:delta-E}):
\begin{equation}
  S_0 = \frac{{\cal A}}{4 l_{\rm P}^2} 
    + O\biggl( \frac{{\cal A}^q}{l_{\rm P}^{2q}};\, q < 1 \biggr),
\label{eq:S_0}
\end{equation}
i.e.\ the states that do not have a field/string theoretic excitation 
on the semiclassical black hole background and in which the stretched 
horizon, located at $r = 2Ml_{\rm P}^2 + O(1/M) \equiv r_{\rm s}$, 
is not excited.

Labeling these exponentially many states by $k$, which we call the 
{\it vacuum index}, basis states for the general microstates of a black 
hole of mass $M$ (within the uncertainty $\varDelta M$) can be given by
\begin{equation}
  \ket{\Psi_{\bar{a}\, a\, a_{\rm far}; k}(M)} 
  \approx \ket{\psi_{\bar{a} a; k}(M)} \ket{\phi_{a_{\rm far}}(M)}.
\label{eq:states}
\end{equation}
Here, $\bar{a}$, $a$, and $a_{\rm far}$ label the excitations of 
the stretched horizon, in the zone (i.e.\ the region within the 
gravitational potential barrier defined, e.g., as $r \leq R_{\rm Z} 
\equiv 3M l_{\rm P}^2$), and outside the zone ($r > R_{\rm Z}$), 
respectively, and $\ket{\psi_{\bar{a} a; k}(M)}$ and 
$\ket{\phi_{a_{\rm far}}(M)}$ are black hole and exterior 
states.  (Here, we have used the fact that $k$ can be regarded 
as being mostly in $r \leq R_{\rm Z}$; see later.)  As we have 
argued, the index $k$ runs over $1, \cdots, e^{S_0}$ for the 
vacuum states $\bar{a} = a = a_{\rm far} =0$.  In general, the 
range for $k$ depends on $\bar{a}$ and $a$, but its dependence is 
higher order in $l_{\rm P}^2/{\cal A}$ so we mostly ignore it. 
This small dependence, however, becomes relevant when we discuss 
negative energy excitations associated with Hawking emission.

Excitations here are defined as fluctuations with respect to a 
fixed background, so their energies as well as entropies can be 
either positive or negative, although their signs must be the 
same.  As discussed in Refs.~\cite{'tHooft:1993gx,Nomura:2013lia}, 
the contribution of the excitations to the total entropy is subdominant 
in $l_{\rm P}^2/{\cal A}$.  The total entropy in the near black hole 
region, $r \leq R_{\rm Z}$, is thus given by $S = {\cal A}/4 l_{\rm P}^2$ 
at the leading order.

The fact that all the independent microstates with different $k$ 
lead to the same geometry suggests that the semiclassical picture 
is obtained after coarse-graining the degrees of freedom represented 
by this index, the vacuum degrees of freedom~\cite{Nomura:2014yka}. 
According to this picture, the black hole vacuum state in the 
semiclassical description is given by the density matrix
\begin{equation}
  \rho_0(M) = \frac{1}{e^{S_0}} \sum_{k=1}^{e^{S_0}} 
    \ket{\Psi_{\bar{a}=a=a_{\rm far}=0; k}(M)} 
    \bra{\Psi_{\bar{a}=a=a_{\rm far}=0; k}(M)}.
\label{eq:rho_0}
\end{equation}
To obtain the response of this state to the operators in the semiclassical 
theory, we may trace out the subsystem on which they do not act.  Denoting 
this subsystem by $\bar{C}$, the relevant reduced density matrix is
\begin{equation}
  \tilde{\rho}_0(M) = {\rm Tr}_{\bar{C}}\, \rho_0(M).
\label{eq:rho_0-sc}
\end{equation}
Consistently with our identification of the origin of the Bekenstein-Hawking 
entropy, we assume that this represents the thermal density matrix
\begin{equation}
  \tilde{\rho}_0(M) \approx 
    \frac{e^{-\beta H_{\rm sc}(M)}}{{\rm Tr}\, e^{-\beta H_{\rm sc}(M)}};
\quad
  \beta = \left\{ \begin{array}{ll}
    \frac{1}{T_{\rm H}} & \mbox{for } r \leq R_{\rm Z}, \\
    +\infty & \mbox{for } r > R_{\rm Z},
    \end{array} \right.
\label{eq:rho_0-therm}
\end{equation}
where $T_{\rm H} = 1/8\pi M l_{\rm P}^2$, and $H_{\rm sc}(M)$ is the 
Hamiltonian of the semiclassical theory.  Here, the expression $\beta 
H_{\rm sc}(M)$ should really be interpreted as $\beta$ times the 
Hamiltonian density integrated over space.

In standard semiclassical field theory, the density matrix of 
Eq.~(\ref{eq:rho_0-therm}) is obtained as a reduced density 
matrix by tracing out the region within the horizon in the {\it unique} 
global black hole vacuum state.  Our view is that this density matrix 
is obtained from a mixed state of exponentially many pure states, 
arising from the coarse-graining in Eq.~(\ref{eq:rho_0}).  We stress 
that the information in the vacuum index $k$ is invisible in the 
semiclassical theory as it is already coarse-grained {\it to obtain} 
the theory; in particular, the dynamics of the vacuum degrees of 
freedom cannot be described in terms of $H_{\rm sc}(M)$.

The expression in Eq.~(\ref{eq:rho_0-therm}) suggests that the spatial 
distribution of the information about $k$ follows the thermal entropy 
calculated using the local temperature:
\begin{equation}
  T(r) \simeq \left\{ \begin{array}{ll} 
    \frac{T_{\rm H}}{\sqrt{1-\frac{2Ml_{\rm P}^2}{r}}} 
      & \mbox{for } r \leq R_{\rm Z}, \\
    0 & \mbox{for } r > R_{\rm Z}.
    \end{array} \right.
\label{eq:T_local}
\end{equation}
In particular, the region around the edge of the zone, $r \leq R_{\rm Z}$ 
and $r - 2Ml_{\rm P}^2 \,\slashed{\ll}\, Ml_{\rm P}^2$, contains 
$O(1)$ bits of information about $k$.

Semiclassical operators in the zone act nontrivially on both $a$ 
and $k$ indices; otherwise the maximal mixture in Eq.~(\ref{eq:rho_0}) 
is not compatible with the thermality in Eq.~(\ref{eq:rho_0-therm}). 
Since the thermal nature is prominent only for modes whose energies 
measured in the asymptotic region are
\begin{equation}
  \omega \lesssim T_{\rm H},
\label{eq:IR-modes}
\end{equation}
this feature is significant only for such infrared modes.  For 
operators with Eq.~(\ref{eq:IR-modes}), their actions on microstates 
can be complicated, although they act on the coarse-grained vacuum 
state of Eq.~(\ref{eq:rho_0}) as if it is the thermal state in 
Eq.~(\ref{eq:rho_0-therm}).

There is a simple physical picture behind this phenomenon of 
``non-decoupling'' of the $a$ and $k$ indices for the infrared modes. 
As viewed from a distance, these modes are ``too soft'' to be resolved 
clearly above the background.  Since the derivation of the semiclassical 
theory involves coarse-graining over microstates in which the energy 
stored in the region $r \lesssim R_{\rm Z}$ has spreads of order 
$\varDelta E \approx 1/M l_{\rm P}^2$, infrared modes with $\omega 
\lesssim T_{\rm H} \approx O(1/M l_{\rm P}^2)$ are not necessarily 
distinguished from ``spacetime fluctuations'' of order $\varDelta E$.

The structure described above leads to the following picture for black 
hole evaporation.%
\footnote{We focus on a single Hawking emission and ignore excitations 
 beyond those directly associated with the emission.  For a more complete 
 discussion, see Ref.~\cite{NSW}.}
Suppose a black hole of mass $M$ is in microstate $k$:
\begin{equation}
  \ket{\Psi_k(M)} = \ket{\psi_k(M)} \ket{\phi_I},
\label{eq:H-before}
\end{equation}
where $\ket{\psi_k(M)}$ is the black hole state, with suppressed 
excitation indices, and $\ket{\phi_I}$ the exterior state.  After 
a timescale of $t \approx O(M l_{\rm P}^2)$, this state evolves due 
to Hawking emission as
\begin{equation}
  \ket{\psi_k(M)} \ket{\phi_I} \rightarrow 
    \sum_{i,a,k'} c^k_{i a k'} \ket{\psi_{a; k'}(M)} \ket{\phi_{I+i}},
\label{eq:H-emission}
\end{equation}
where $\ket{\phi_{I+i}}$ is the state in which newly emitted Hawking 
quanta, labeled by $i$ and having energy $E_i$, are added to the 
appropriately time evolved $\ket{\phi_I}$.  The index $a$ represents 
the fact that the black hole state has negative energy excitations 
of energy $-E_a$ around the edge of the zone, created in connection 
with the Hawking emission; the coefficients $c^k_{i a k'}$ are nonzero 
only if $E_i \approx E_a$ (within the uncertainty).  The negative 
energy excitations then propagate inward, and after a time of order 
$M l_{\rm P}^2 \ln(M l_{\rm P})$ collide with the stretched horizon, 
making the black hole states relax as
\begin{equation}
  \ket{\psi_{a; k'}(M)} \rightarrow 
    \sum_{k_a} d^{a k'}_{k_a} \ket{\psi_{k_a}(M-E_a)}.
\label{eq:H-relax}
\end{equation}
The combination of Eqs.~(\ref{eq:H-emission},~\ref{eq:H-relax}) yields
\begin{equation}
  \ket{\psi_k(M)} \ket{\phi_I} \rightarrow 
    \sum_{i,k_i} \alpha^k_{i k_i} \ket{\psi_{k_i}(M-E_i)} \ket{\phi_{I+i}},
\label{eq:H-process}
\end{equation}
where $\alpha^k_{i k_i} = \sum_{a,k'} c^k_{i a k'} d^{a k'}_{k_i}$, 
and we have used $E_i = E_a$.  This expression shows that information 
in the black hole can be transferred to the radiation state $i$.

It is important that the negative energy excitations in 
Eq.~(\ref{eq:H-emission}) come with {\it negative entropies}, so that 
each of the processes in Eqs.~(\ref{eq:H-emission},~\ref{eq:H-relax}) 
is separately unitary.  Specifically, as $k$ and $i$ run over all the 
possible values with $a$ being fixed, the index $k'$ runs only over 
$1,\cdots,e^{S_0(M-E_a)}$, the dimension of the space spanned by 
$k_a$.  Here, $S_0(M) \equiv 4\pi M^2 l_{\rm P}^2$.  This is an 
example of the non-factorizable nature of the $k$ and $a$ indices 
discussed after Eq.~(\ref{eq:states}).  This structure avoids the 
firewall argument in Ref.~\cite{Almheiri:2013hfa}---unlike what 
is imagined there, the physical Hilbert space is smaller than the 
naive Fock space built on each $k$.

From the semiclassical standpoint, the emission of Eq.~(\ref{eq:H-emission}) 
is viewed as occurring locally around the edge of the zone, which is 
possible because the information about the black hole microstate extends 
into the whole zone region.  In this region, information stored in the 
vacuum state, $k$, is transferred into that in modes $a_{\rm far} \neq 0$, 
which have clear identities over the background spacetime.  Because 
of energy conservation, this process is accompanied by the creation 
of ingoing negative energy excitations.  Unlike standard pair creation, 
however, these excitations are {\it not} (maximally) entangled with 
the emitted Hawking quanta.

The discussion here indicates that the purifiers of the emitted 
Hawking quanta are microstates which semiclassical theory describes 
as a vacuum.  Unlike what was considered in Ref.~\cite{Almheiri:2012rt}, 
Hawking quanta are not modes associated solely with one of the Rindler 
wedges in the near horizon approximation ($b$ modes in the notation 
of Ref.~\cite{Almheiri:2012rt}) nor outgoing Minkowski modes ($a$ 
modes), which would appear to have high energies for infalling observers. 
This allows for avoiding the entropy~\cite{Almheiri:2012rt} and 
typicality~\cite{Marolf:2013dba} arguments for firewalls.  Note that 
physics described here need not introduce nonlocality in low energy 
field theory; it can still respect causality in $r > r_{\rm s}$.

We emphasize that the vacuum degrees of freedom play {\it dual} roles. 
While they represent how the semiclassical spacetime is composed at 
the microscopic level, they also appear as thermal radiation when 
probed in the semiclassical theory.  In fact, these degrees of freedom 
are neither spacetime nor matter/radiation.  In particular, their 
detailed dynamics cannot be treated in semiclassical theory.

The above understanding of Hawking emission clarifies why the 
semiclassical calculation of Ref.~\cite{Hawking:1976ra} finds an 
apparent violation of unitarity.  At the microscopic level, formation 
and evaporation of a black hole involve the vacuum degrees of freedom. 
Since semiclassical theory is incapable of describing their microscopic 
dynamics, the description of black hole evolution in semiclassical 
theory is necessarily non-unitary.

A similar analysis can also be performed for black hole 
mining~\cite{Unruh:1982ic,Brown:2012un}.  See Ref.~\cite{NSW} 
for details.

\section{Infalling Description}

Suppose we drop an object into a black hole.  In a distant reference 
frame, the semiclassical description of the object (in terms of $a$ 
and $a_{\rm far}$) is applicable only until it hits the stretched 
horizon, after which it is represented as excitations of the stretched 
horizon (in terms of $\bar{a}$).  The information about the fallen 
object will then stay there, at least, for the scrambling time of order 
$M l_{\rm P}^2 \ln(M l_{\rm P})$~\cite{Hayden:2007cs} before being 
transferred to $k$.  On the other hand, the equivalence principle 
says that the falling object does not feel anything special when it 
crosses the horizon.  How can these two pictures be consistent?

The idea of complementarity is that the infalling object is still 
described using low energy language after it crosses the Schwarzschild 
horizon by making an appropriate reference frame change.  Here we 
consider a class of reference frames which reveal the spacetime 
structure near the Schwarzschild horizon in the clearest form. 
We call them {\it infalling reference frames}.

Let the spatial origin $p_0$ of a reference frame follow a timelike 
geodesic released from rest at $r = r_0$, with $r_0 - 2 M l_{\rm P}^2 
\gtrsim M l_{\rm P}^2$.  According to complementarity, the system 
described in this reference frame does not have a (hot) stretched 
horizon at the location of the Schwarzschild horizon when $p_0$ 
crosses it; the region around $p_0$ appears approximately flat 
up to small curvature effects.

In this description, a ``horizon'' signaling the breakdown of the 
semiclassical description is expected to appear in the past-directed 
and inward directions from $p_0$.  In analogy with the case of a 
distant frame description, we denote basis states for the general 
microstates as
\begin{equation}
  \ket{\Psi_{\bar{\alpha}\, \alpha\, \alpha_{\rm far}; \kappa}(M)},
\label{eq:states-falling}
\end{equation}
where $\bar{\alpha}$ labels the excitations of the ``horizon,'' and 
$\alpha$, and $\alpha_{\rm far}$ the semiclassical excitations near and 
far from the black hole, respectively; $\kappa$ is the vacuum index.

The complementarity transformation provides a map between the states 
in Eq.~(\ref{eq:states}) and those in Eq.~(\ref{eq:states-falling}). 
While the general form of this transformation can be complicated, 
we may consider, based on the analysis of an infalling object, that 
a portion of the $\alpha$ index representing interior excitations 
is transformed into the $\bar{a}$ index (and vice versa).  Note that 
the amount of information needed to reconstruct the interior (in 
the semiclassical sense) is much smaller than the Bekenstein-Hawking 
entropy---the logarithm of the dimension of the relevant Hilbert 
space is of order $({\cal A}/l_{\rm P}^2)^q$ with $q < 1$.

Where are the $\kappa$ degrees of freedom located?  We expect that 
most are in the region close to the ``horizon''; in particular, the 
number of $\kappa$ degrees of freedom within a distance sufficiently 
smaller than $M l_{\rm P}^2$ from $p_0$ is of $O(1)$, since the time 
and length scales characterizing local deviations from Minkowski 
space are of order $M l_{\rm P}^2$ there.  This invites a question:\ 
how can this picture be consistent with that in the distant reference 
frame, which has a very different spacetime distribution of the 
vacuum degrees of freedom?

To see a nontrivial consistency between the two pictures, consider 
detectors hovering at a constant $r$ with $r - 2Ml_{\rm P}^2 \ll M 
l_{\rm P}^2$.  In a distant description, the spatial density of the 
microscopic information in $k$ is large there, so that these detectors 
can be used for black hole mining.  The rate of extracting information, 
however, is still of order one qubit per Schwarzschild time $t \approx 
O(M l_{\rm P}^2)$ {\it per channel}~\cite{Brown:2012un}---the acceleration 
of information extraction occurs not because of a higher rate in each 
channel but because of an increased number of available channels. 
This implies that each single detector, which we define to act on 
a single channel, ``clicks'' once per $t \approx O(M l_{\rm P}^2)$.

In an infalling reference frame, the density of the microscopic 
information in $\kappa$ is small at the detector location, at least 
when $p_0$ passes nearby.  The rate of extracting information thus 
cannot be much faster than $1/M l_{\rm P}^2$ around $p_0$, reflecting 
the fact that the spacetime appears approximately flat there.  This, 
however, is still consistent with the distant description.  By adopting 
the near-horizon Rindler approximation, one can show that when viewed 
from the infalling reference frame, the detector clicks only once 
in each time/space interval of
\begin{equation}
  \varDelta T \approx \varDelta Z \approx O(M l_{\rm P}^2),
\label{eq:click-TZ}
\end{equation}
around $p_0$~\cite{NSW}.  This is what we expect from the equivalence 
principle:\ the spacetime appears flat up to curvature effects with 
lengthscale $M l_{\rm P}^2$.  While the detector clicks of order 
$\ln(M l_{\rm P})$ times within the causal patch of the infalling 
frame, these clicks occur at distances of order $M l_{\rm P}^2$ away 
from $p_0$, where we expect a higher density of $\kappa$ degrees 
of freedom.

The two descriptions are thus consistent.  It is striking that the 
microscopic information about a black hole exhibits this level of 
reference frame dependence, a phenomenon we refer to as {\it extreme 
relativeness}.

\section{Other Reference Frames}

We now discuss a reference frame whose origin follows a timelike 
geodesic released from rest at $r = r_0$, where $r_0$ is {\it close 
to} the Schwarzschild horizon, $r_0 - 2 M l_{\rm P}^2 \ll M l_{\rm P}^2$. 
In the case of $r_0 - 2 M l_{\rm P}^2 \gtrsim M l_{\rm P}^2$, we 
found that the detector-click time/length scales are given by 
Eq.~(\ref{eq:click-TZ}), despite the fact that the detector clicks 
at a much higher rate in its own frame.  Technically, this was due 
to a huge relative boost between $p_0$ and the detector when they 
approach.  Here, however, the relevant boost is not as large, and 
the detector-click time/length scales appear as
\begin{equation}
  \varDelta T \approx \varDelta Z \ll M l_{\rm P}^2.
\label{eq:click-TZ-2}
\end{equation}

Since each detector click extracts an $O(1)$ amount of information 
from spacetime, which we expect not to occur in Minkowski space, this 
implies that the spacetime as viewed from this reference frame is not 
approximately Minkowski over the lengthscale $M l_{\rm P}^2$ when 
$p_0$ crosses the Schwarzschild horizon.  We interpret this to mean 
that in this reference frame, the ``horizon'' is at a distance of order 
$\varDelta Z$ away from $p_0$, so that detector clicks occur near or 
``on'' this surface.  Since we expect that the microscopic information 
is located near and on the ``horizon,'' there is no inconsistency 
for the clicks to extract information from the black hole.

One might worry that in this reference frame, spacetime near the 
Schwarzschild horizon does not appear large, $\approx O(M l_{\rm P}^2)$, 
nearly flat space.  However, the {\it existence} of an infalling 
reference frame discussed before ensures that an infalling physical 
observer sees a large black hole interior.  The analysis here simply 
says that the spacetime around the Schwarzschild horizon is not 
always {\it described} as a large nearly flat region, even in 
reference frames falling freely into the black hole.

We finally discuss (non-)relations of black hole mining and the 
Unruh effect~\cite{Unruh:1976db} in Minkowski space.  It is often 
thought that these two reveal the same physics, which would mean 
the existence of a ``horizon'' in an {\it inertial} frame description 
of Minkowski space.  This is, however, not true.  Since the equivalence 
principle can make a statement only about a point at a given moment 
in a given reference frame, while a system in quantum mechanics 
is specified by a state which encodes global information on the 
equal-time hypersurface, there is no reason that physics of the 
two systems must be similar beyond a point in space.  In particular, 
the inertial frame description of Minkowski space does not have 
a ``horizon,'' so a detector reacts very differently to blueshifted 
Hawking radiation and Unruh radiation in Minkowski space---it extracts 
microscopic information about spacetime in the former case, while 
it does not in the latter.  The relation between quantum mechanics 
and the equivalence principle seems subtle, but they are consistent.

\section*{Acknowledgments}

We would like to thank Raphael Bousso, Ben Freivogel, Daniel Harlow, 
Juan Maldacena, Donald Marolf, Joseph Polchinski, Douglas Stanford, 
Jaime Varela, Erik Verlinde, Herman Verlinde, and I-Sheng Yang for 
various conversations during our exploration of this subject.  Y.N. 
thanks the Aspen Center for Physics and the National Science Foundation 
(NSF) Grant \#1066293 for hospitality during his visit in which a part 
of this work was carried out.  F.S. thanks the Department of Energy 
(DOE) National Nuclear Security Administration Stewardship Science 
Graduate Fellowship for financial support.  This work was supported 
in part by the Director, Office of Science, Office of High Energy and 
Nuclear Physics, of the U.S.\ DOE under Contract DE-AC02-05CH11231, 
and in part by the NSF under grant PHY-1214644.

\end{document}